# Pinning-depinning mechanism of the contact line during evaporation on chemically patterned surfaces: A lattice Boltzmann study

Qing Li,*[†,‡] P. Zhou,[†] and H. J. Yan[†]

[†]School of Energy Science and Engineering, Central South University, Changsha 410083, China

[‡]Computational Earth Science Group, Los Alamos National Laboratory, Los Alamos, NM 87545, USA

In this paper, the pinning and depinning mechanism of the contact line during droplet evaporation on chemically stripe-patterned surfaces is numerically investigated using a thermal multiphase lattice Boltzmann (LB) model with liquid-vapor phase change. A local force balance in the context of diffuse interfaces is introduced to explain the equilibrium states of droplets on chemically patterned surfaces. It is shown that, when the contact line is pinned on a hydrophobic-hydrophilic boundary, different contact angles can be interpreted as the variation of the length of the contact line occupied by each component. The stick-slip-jump behavior of evaporating droplets on chemically patterned surfaces is well captured by the LB simulations. Particularly, a slow movement of the contact line is clearly observed during the stick (pinning) mode, which shows that the pinning of the contact line during droplet evaporation on chemically stripe-patterned surfaces is actually a dynamic pinning process and the dynamic equilibrium is achieved by the self-adjustment of the contact lines occupied by each component. Moreover, it is shown that, when the surface tension varies with the temperature, the Marangoni effect has an important influence on the depinning of the contact line, which occurs when the horizontal component (towards the center of the droplet) of the force caused by the Marangoni stress overcomes the unbalanced Young's force towards outside.

*Corresponding author: qingli@csu.edu.cn

# 1. Introduction

Droplet evaporation is a fundamental phenomenon in nature [1] and plays a vital role in various fields of natural science and engineering such as spray cooling [2], inkjet printing [3], coating [4] and chip manufacturing [5, 6]. Understanding the underlying physical mechanisms is of great importance in utilizing droplet evaporation-based processes in these applications. To date, numerous studies have been performed to investigate droplet evaporation on various surfaces. Several comprehensive reviews have been presented, e.g., by Cazabat and Guéna [7], Erbil [1], and Kovalchuk *et al*. [8]. In the pioneering work of Picknett and Bexon [9], they identified two modes in the evaporation of droplets on smooth surfaces, namely the constant contact radius (CCR) mode and the constant contact angle (CCA) mode. In the CCR mode, the contact line is pinned on the solid surface while the contact angle decreases. In the CCA mode, the contact angle remains constant while the contact line recedes towards the center of the droplet.

Droplet evaporation on solid surfaces involves a number of complex transport mechanisms [1, 10]. Moreover, the surface morphology and chemical composition also affect the behavior of droplet evaporation significantly. In spite of the fact that tremendous research efforts have been made, many aspects of droplet evaporation are still not well understood. One of the aspects is the mechanism of the pinning and depinning of the contact line during evaporation. Recently, much attention has been paid to a pinning-depinning (or stick-slip) phenomenon and CCR-CCA transition during droplet evaporation on structurally rough surfaces [11-20] and chemically heterogeneous surfaces [21-23]. For example, Orejon *et al*. [12] investigated the dynamics of the three-phase contact line of evaporating droplets on different surfaces with varying substrate hydrophobicity. They proposed that the depinning of contact line on rough surfaces occurs when the unbalanced Young's force is large enough to overcome an intrinsic energy barrier arising from surface roughness. Chen *et al*. [13] studied the contact line dynamics during droplet evaporation on micro-structured surfaces. In their work, the pinning force and the depinning force were formulated.

Several studies of the pinning-depinning phenomenon during droplet evaporation on chemically heterogeneous surfaces (patterned with hydrophobic and hydrophilic stripes) have also been reported [21-23]. Wang and Wu [21] firstly performed molecular dynamics simulations for the evaporation of nano-droplets on chemically stripe-patterned surfaces. Their simulations were carried out under the constant-temperature condition and the evaporation was implemented by extracting the liquid molecules from the droplet surface. The pinning-depinning phenomenon (CCR-CCA transition) was observed during the evaporation on the surfaces with relatively large stripe widths. Jansen *et al*. [22] experimentally studied the evaporation of elongated droplets on chemically stripe-patterned surfaces. It was found that, when the droplets become spherical, they evaporate in a stick-slip fashion in the perpendicular side. In a very recent paper, Zhang *et al*. [23] pointed out that the pinning during droplet evaporation on chemically heterogeneous surfaces should be interpreted as a drastic slowdown of the contact line dynamics. In Zhang *et al*.'s molecular dynamics study, the evaporation was also triggered by removing molecules from the system at a given rate.

Despite the existing studies, the mechanism of the pinning-depinning phenomenon during droplet evaporation on chemically heterogeneous surfaces is still not very clear. In this work, we aim at providing an understanding of the pinning-depinning behavior of evaporating droplets on chemically stripe-patterned surfaces from the lattice Boltzmann (LB) method [24-29], which can be viewed as a minimal form of the Boltzmann equation in the kinetic theory [24]. In the LB method, the interface between different phases can arise, deform and migrate naturally, without resorting to any techniques to track or capture the interface [29, 30]. In terms of the modeling of interfacial phenomena, the LB method can be classified into the diffuse interface method [31] because the interface in LB simulations is usually a diffuse interface that is of finite thickness (around 4-5 grids).

In the literature, there have been some LB studies about droplet dynamics on chemically patterned surfaces under the isothermal (constant-temperature) condition [32-38]. We also notice that Ledesma-Aguilar *et al.* [39] have conducted an LB study of droplet evaporation using a phase-field multiphase LB model. In their work the LB simulations were carried out under the isothermal condition and the evaporation was driven by concentration gradients.

In the present work, a hybrid thermal multiphase LB model with the liquid-vapor phase change being taken into account is employed to study the droplet evaporation on chemically stripe-patterned surfaces. The model was developed in our previous studies [40, 41] and has been validated by mimicking the three boiling stages in pool boiling as well as the boiling curve [40]. In addition, its capability has also been demonstrated by investigating the self-propelled motion of Leidenfrost droplets on ratchet surfaces [41]. The rest of the present paper is organized as follows. Section 2 briefly introduces the thermal multiphase LB model. Section 3 is devoted to the numerical results and the corresponding discussions. Finally, Section 4 summarizes the present paper.

## 2. Numerical model

The LB method [24, 25, 27-29] historically originated from the lattice gas automata method [42]. Later He and Luo [43] demonstrated that the LB equation can be rigorously derived from the Boltzmann equation in the kinetic theory. Thus the LB method can be regarded as a mesoscopic approach that simulates fluid flows by solving the discrete Boltzmann equation with a certain collision operator, such as the Bhatnagar-Gross-Krook collision operator [44] and the Multiple-Relaxation-Time (MRT) collision operator [45, 46]. The macroscopic averaged properties are obtained by accumulating the density distribution function. Generally, the LB equation, which governs the evolution of the density distribution function, can be written as follows:

$$f_\alpha \left( \mathbf{x} + \mathbf{e}_\alpha \delta_t, t + \delta_t \right) = f_\alpha \left( \mathbf{x}, t \right) - \widehat{\Lambda}_{\alpha\beta} \left( f_\beta - f_\beta^{eq} \right) \Big|_{(\mathbf{x}, t)} + \delta_t F'_\alpha \left( \mathbf{x}, t \right), \tag{1}$$

where $f_\alpha$ is the density distribution function, $f_\alpha^{eq}$ is the equilibrium distribution, $t$ is the time, $\mathbf{x}$

is the spatial position, $\mathbf{e}_\alpha$ is the discrete velocity along the $\alpha$ th direction, $\delta_t$ is the time step, $F'_\alpha$ represents the forcing term in the velocity space, and $\hat{\Lambda}_{\alpha\beta} = \left(\mathbf{M}^{-1}\Lambda\mathbf{M}\right)_{\alpha\beta}$ is the collision matrix, in which $\Lambda$ is a diagonal matrix and $\mathbf{M}$ is an orthogonal transformation matrix [45]. In the LB method, the equilibrium density distribution function is usually formulated as [44]

$$f_\alpha^{eq} = \omega_\alpha \rho \left[ 1 + \frac{\mathbf{e}_\alpha \cdot \mathbf{v}}{c_s^2} + \frac{\mathbf{v}\mathbf{v} : \left(\mathbf{e}_\alpha \mathbf{e}_\alpha - c_s^2 \mathbf{I}\right)}{2c_s^4} \right], \quad (2)$$

where $\mathbf{I}$ is the unit tensor, $c_s$ is the lattice sound speed, and $\omega_\alpha$ are the weights. The macroscopic density and velocity are calculated by

$$\rho = \sum_\alpha f_\alpha, \quad \rho \mathbf{v} = \sum_\alpha \mathbf{e}_\alpha f_\alpha + \frac{\delta_t}{2}\mathbf{F}, \quad (3)$$

where $\mathbf{F}$ is the force acting on the system. With the aid of the transformation matrix $\mathbf{M}$, the right-hand side of Eq. (1) can be carried out in the moment space

$$\mathbf{m}^* = \mathbf{m} - \Lambda\left(\mathbf{m} - \mathbf{m}^{eq}\right) + \delta_t \left(\mathbf{I} - \frac{\Lambda}{2}\right)\mathbf{S}, \quad (4)$$

where $\mathbf{m} = \mathbf{M}\mathbf{f}$, $\mathbf{m}^{eq} = \mathbf{M}\mathbf{f}^{eq}$, and $\mathbf{S}$ is the forcing term in the moment space. Then the left-hand side of Eq. (1) can be obtained via $\mathbf{f} = \mathbf{M}^{-1}\mathbf{m}^*$, where $\mathbf{f} = \left(f_0, f_1, \cdots, f_{n-1}\right)^{\mathrm{T}}$.

For single-component multiphase flows, the pseudopotential interaction force proposed by Shan and Chen [47, 48] is employed, which has been applied in a variety of fields with great success [29] and is given by

$$\mathbf{F} = -G\psi(\mathbf{x}) \sum_\alpha w_\alpha \psi(\mathbf{x} + \mathbf{e}_\alpha \delta_t)\mathbf{e}_\alpha, \quad (5)$$

where $\psi$ is the pseudopotential, $G$ is the interaction strength, and $w_\alpha$ are the weights. The force $\mathbf{F}$ is incorporated into Eq. (4) through the forcing term $\mathbf{S}$ [40, 41]. To reproduce a non-ideal equation of state, the pseudopotential $\psi$ is taken as $\psi(\mathbf{x}) = \sqrt{2\left(p_{eos} - \rho c_s^2\right)/Gc^2}$ [49], where $p_{eos}$ is a non-ideal equation of state and $c = 1$ is the lattice constant. For the diffuse interface modeling of multiphase flows, the temperature equation can be derived from the local balance law for entropy [31], and is given by

$$\rho c_v \frac{DT}{Dt} = \nabla \cdot (\lambda \nabla T) - T \left( \frac{\partial p_{eos}}{\partial T} \right)_\rho \nabla \cdot \mathbf{v} . \tag{6}$$

The temperature equation is numerically solved with the fourth-order Runge-Kutta scheme for time discretization and the isotropic central scheme for spatial discretization [40].

The coupling between the modeling of the fluid flow and the solver for the temperature field is established via the non-ideal equation of state $p_{eos} = p_{eos}(\rho, T)$, through which the pseudopotential $\psi(\mathbf{x}) = \sqrt{2(p_{eos} - \rho c_s^2)/Gc^2}$ in the interaction force, Eq. (5), is linked to the temperature field. Then the liquid-vapor phase change can be driven by the temperature field via the equation of state. In the present work, we adopt the Peng-Robinson equation of state [49]. The saturation temperature is set to $T_{sat} = 0.86 T_c$ ($T_c$ is the critical temperature), which corresponds to the liquid density $\rho_L \approx 6.5$ and the vapor density $\rho_V \approx 0.38$. The liquid-vapor density ratio is about 17. Here it is worth mentioning that the quantities in LB simulations are usually based on the lattice units [37, 39].

## 3. Numerical results and discussions

### 3.1 The equilibrium contact angles and the force balance without evaporation

The chemically heterogeneous surfaces patterned with hydrophobic and hydrophilic stripes are illustrated in Fig. 1, where "A" and "B" represent the hydrophilic and hydrophobic stripes, respectively. The width of the hydrophilic stripe is equal to that of the hydrophobic stripe. In addition, we only study the cases in which the center of mass of the droplet is on the middle of a hydrophilic stripe (see Fig. 1). Similar to some of the previous studies involving chemically stripe-patterned surfaces [21, 23, 33, 36], the present work also investigates a two-dimensional situation so as to retain the essential features of the stick-slip dynamics.

The equilibrium contact angles on homogeneous surfaces are implemented through the geometric formulation proposed by Ding and Spelt [50]. Using the second-order central difference scheme, the geometric formulation can be given by $\rho_{i,0} = \rho_{i,2} + \tan(90° - \theta_e^a)|\rho_{i+1,1} - \rho_{i-1,1}|$, where $\theta_e^a$ is an analytically prescribed static contact angle and $\rho_{i,0}$ represents the density at the ghost layer $(i, 0)$ beneath the solid wall [50], in which the first index denotes the coordinate along the solid wall while the second index denotes the coordinate normal to the solid wall. For $\theta_e^a = 60°$ and $120°$, we numerically obtained $\theta_e \approx 63.3°$ and $119°$ from our LB simulations, which serve as the equilibrium contact angles $\theta_{A,e}$ and $\theta_{B,e}$ for the hydrophilic and hydrophobic stripes used in the present study.

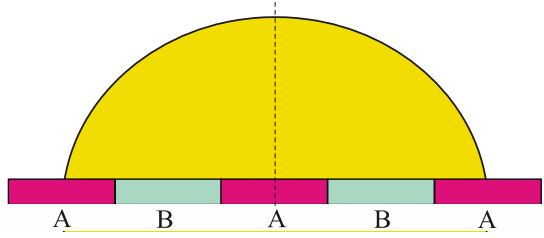

**Fig. 1** A schematic diagram of chemically stripe-patterned surfaces. "A" and "B" represent the hydrophilic and hydrophobic stripes, respectively.

Our simulations are carried out in a two-dimensional rectangular domain $L_x \times L_y = 300 \times 150$ (lattice units). The kinematic viscosity is taken as $\nu = 0.07$ and the thermal diffusivity $\chi = \lambda/\rho$, where $\lambda$ is the thermal conductivity, is set to 0.15. The Zou-He boundary scheme [51] is applied at the solid wall. The open boundary condition is employed at the top boundary and the periodic boundary condition is applied in the $x$ direction. The interface thickness in our simulations is around 5 l.u. (l.u. represents lattice units). Three cases are considered to investigate the effect of the stripe width. To ensure that the stripe width is larger than the interface thickness, it is set to 9, 11, and 17 l.u. The corresponding chemically stripe-patterned surfaces are named W9, W11, and W17, respectively. Initially, a droplet of radius $r = 40$ (l.u.) is deposited on the middle of the central hydrophilic stripe of the surface. After the spreading and receding processes, the droplet gradually reaches its equilibrium state. The equilibrium states of the droplets on the W9, W11, and W17 surfaces are displayed in Figs. 2(a), 2(b), and 2(c), respectively. Local details of the liquid-vapor interfaces are presented in Figs. 2(d)-2(f). It should be noted the computational domain is larger than the domain shown in Fig. 2. The corresponding equilibrium contact angles are $87.7^\circ$, $101.3^\circ$, and $111.8^\circ$, respectively.

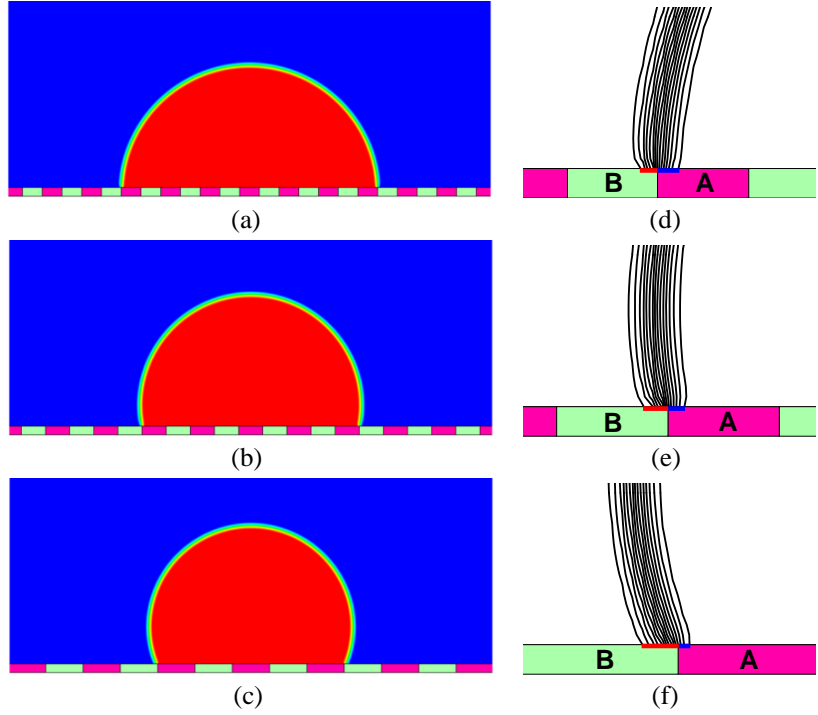

**Fig. 2** Left column, the equilibrium states on the W9, W11, and W17 surfaces (from top to bottom). Right column, local details of the liquid-vapor interfaces ($0.5 \leq \rho \leq 5.5$), where the solid blue line denotes $L_A$ and the solid red line denotes $L_B$. The equilibrium contact angles on the pure A and B surfaces are $63.3^\circ$ and $119^\circ$, respectively.

For a heterogeneous surface consisting of two components A and B, Cassie [52] has proposed the following equation:

$$\cos\theta = f_A \cos\theta_{A,e} + f_B \cos\theta_{B,e}, \tag{7}$$

where $\theta$ is the equilibrium contact angle on the heterogeneous surface; $\theta_{A,e}$ and $\theta_{B,e}$ are the equilibrium contact angles on the pure A and B surfaces, respectively; $f_A$ and $f_B$ ($f_A + f_B = 1$) are the area fractions of the components A and B, respectively, which are obtained according to the liquid-solid interface beneath the droplet. However, it has been widely suggested that the contact angle on a heterogeneous surface is determined by the interactions occur at the three-phase contact line alone. For example, Extrand [53] showed that the three-phase structure at the contact line, not the liquid-solid interfacial contact area, controls the wetting of heterogeneous surfaces. Gao and McCarthy [54] pointed out that Cassie's equation is valid to the extent that the structure of the contact area reflects the ground state energies of contact lines and the fraction of the surface probed by the contact line is important in affecting the contact angle.

In this study we would like to provide an understanding of the equilibrium contact angles in the context of diffuse interface modeling. By considering its thickness, the liquid-vapor interface can be denoted by a cluster of isodensity lines as shown in Figs. 2(d)-2(f). With the concept of a cluster of isodensity lines, we can see that some of the isodensity lines are sitting on an A stripe while others are sitting on a B stripe. For each case, it can be seen that the *contact line* between the liquid-vapor interface and the solid surface is located on a hydrophobic-hydrophilic boundary. However, when focusing on the length of the contact line, we can observe that the length ratios (length of the contact line occupied by each component per total length) vary in the three cases.

It should be noted that in three-dimensional (3D) space the "length" of the contact line usually denotes the perimeter length of the contact line. In the present study, two-dimensional simulations are performed and the length of the contact line describes the thickness of the contact line in 3D space. According to Fig. 2, it can be seen that the length of the contact line occupied by the component B increases from Fig. 2(d) to Fig. 2(f). In what follows we shall show that such a phenomenon can be explained based on a *local force balance*, which would also be a basis for understanding the pinning-depinning phenomenon during droplet evaporation on chemically stripe-patterned surfaces.

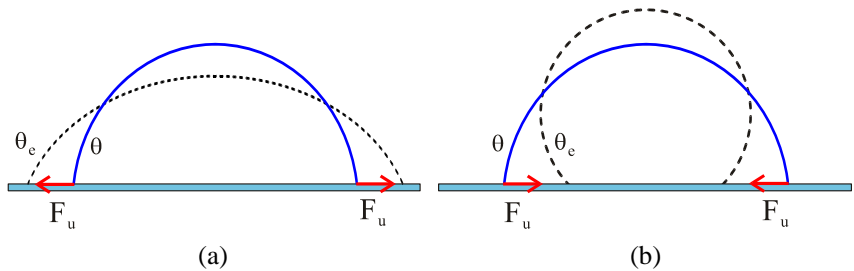

**Fig. 3** The unbalanced Young's force: (a) $\theta > \theta_e$ and (b) $\theta < \theta_e$.

When the contact angle deviates from the equilibrium contact angle, an unbalanced Young's force can be defined [55]: $F_u = \gamma(\cos\theta - \cos\theta_e)$ (per unit length), where $\gamma$ is the liquid-vapor surface tension, $\theta_e$ is the equilibrium contact angle, and $\theta$ is an arbitrary contact angle. The unbalanced Young's force is illustrated in Fig. 3. According to Fig. 3, the isodensity lines sitting on the B stripe in Fig. 2 yield an unbalanced Young's force towards the center of the droplet, while the isodensity lines sitting on the A stripe generate an unbalanced Young's force towards outside. Hence the local force balance in Figs. 2(d)-2(f) along the horizontal direction can be formulated as follows:

$$L_B \gamma (\cos\theta - \cos\theta_{B,e}) = -L_A \gamma (\cos\theta - \cos\theta_{A,e}), \tag{8}$$

where $L_A$ and $L_B$ are the lengths of the contact line occupied by the components A and B, respectively. Since $L_A + L_B = L$, in which $L$ is the total length of the contact line, the following equation can be obtained:

$$\cos\theta = l_A \cos\theta_{A,e} + l_B \cos\theta_{B,e}, \tag{9}$$

where $l_A = L_A/L$ and $l_B = L_B/L$ are the length ratios. Obviously, the above equation takes the same form as the Cassie equation given by Eq. (7). However, the coefficients in front of $\cos\theta_{A,e}$ and $\cos\theta_{B,e}$ are defined differently. Here the equation is derived from the local force balance. Similar equations have been suggested by Park *et al.* [56] and Zhang *et al.* [23]. It is noticed that Park *et al.*'s equation was derived for spherical droplets through calculating the work done for moving the contact line by an infinitesimal distance. In their equation, the coefficients are the length ratios of each components on the perimeter of the contact line. The thickness of the contact line was not taken into account in their derivation. Our equation, Eq. (9), highlights the thickness of the contact line from a two-dimensional view in the context of diffuse interface modeling. When it is extended to spherical droplets, the corresponding coefficients would be the area ratios of each components on the three-phase contact area (not the liquid-solid contact area).

According to Figs. 2(d)-2(f), the length ratio $l_B$ in the three cases is approximately $0.45$, $0.625$, and $0.8$, respectively. The corresponding equilibrium contact angles estimated from Eq. (9) are about $88.3^\circ$, $98^\circ$, and $107.3^\circ$, which are in good agreement with the above mentioned numerical results. For the W9, W11, and W17 surfaces, the contact lines at the equilibrium state are pinned on a hydrophobic-hydrophilic boundary. Actually, we have also studied the cases in which the stripe width is larger than $17$ l.u. For example, when the stripe width is 35 l.u., the contact line at the equilibrium state is completely located on a hydrophobic stripe and the equilibrium contact angle is equal to $\theta_{B,e}$, which also obeys Eq. (9).

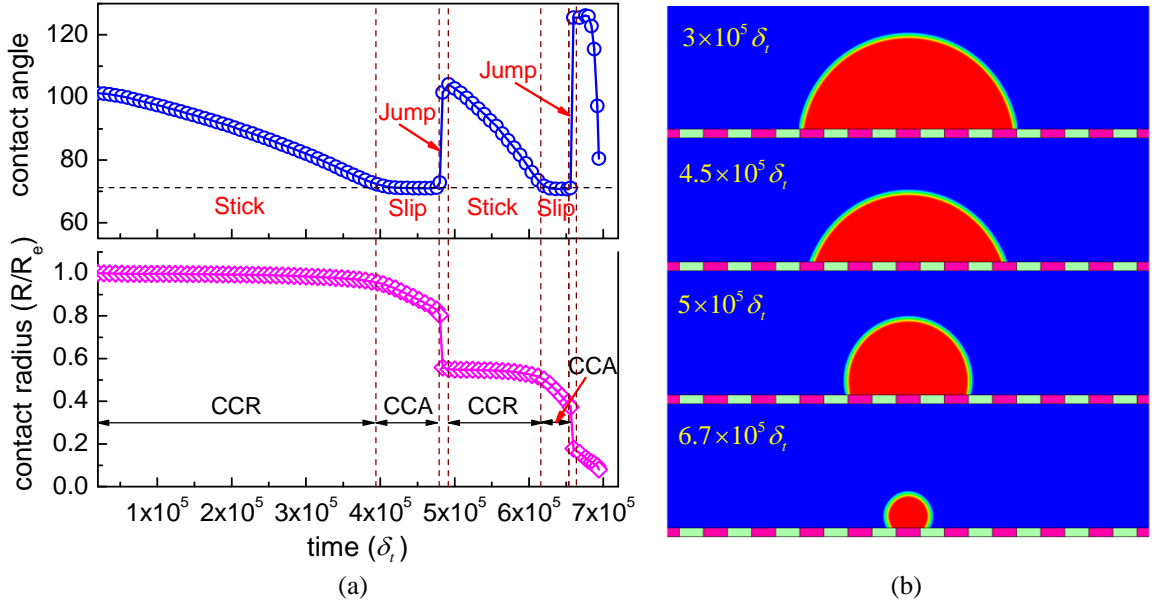

**Fig. 4** Droplet evaporation on the W11 surface. (a) Time evolutions of the contact angle (degrees) and the non-dimensional contact radius; (b) snapshots of the droplet during evaporation.

### 3.2 The stick-slip dynamics and the pinning-depinning mechanism

Now attention turns to the dynamics of evaporating droplets on the aforementioned W9, W11, and W17 surfaces. The temperature solver is added at $t = 20000\delta_t$ with the temperature of the solid wall being set to $T_w = 0.875T_c$. At that time the droplet has reached its equilibrium state in the absence of evaporation. The numerical results of droplet evaporation on the W11 surface are shown in Fig. 4, in which $R$ is the transient contact radius and $R_e$ is the contact radius at $t = 2\times10^4 \delta_t$. Figure 4(a) displays the variations of the contact angle and the non-dimensional contact radius ($R/R_e$) with respect to time. From the figure we can see that the droplet initially evaporates in the CCR mode. The contact line is pinned on a hydrophobic-hydrophilic (B-A) boundary (due to symmetry, only the left contact line is discussed), which can be seen from the top snapshot in Fig. 4(b).

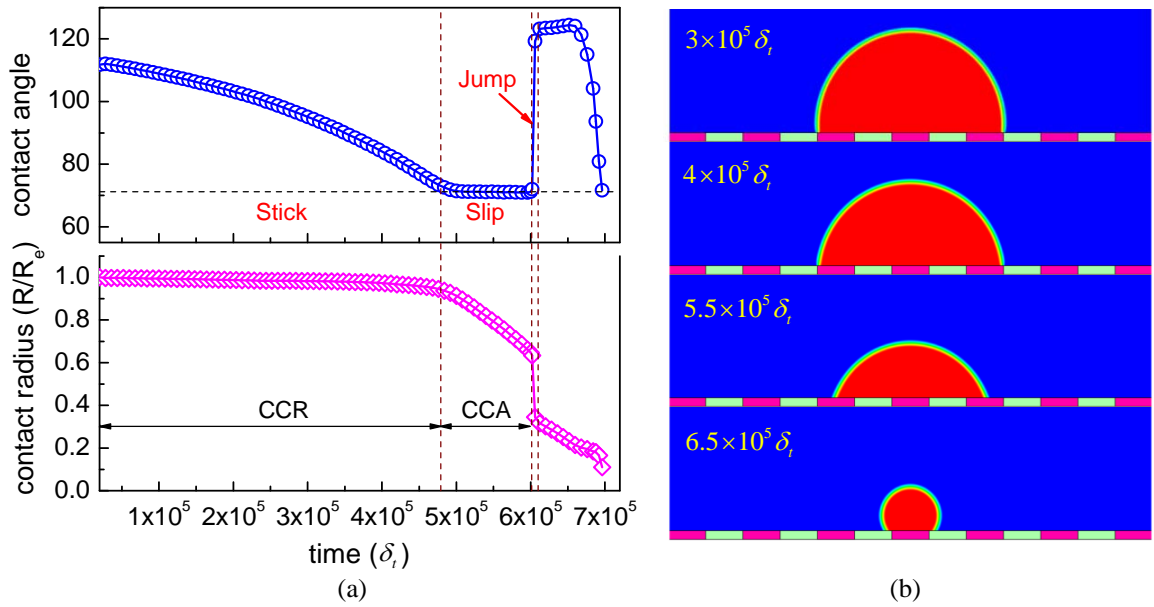

**Fig. 5** Droplet evaporation on the W17 surface. (a) Time evolutions of the contact angle (degrees) and the non-dimensional contact radius; (b) snapshots of the droplet during evaporation.

Subsequently, the evaporation goes to the CCA mode and the droplet starts to slide on the surface as shown in the second snapshot in Fig. 4(b). The CCR-CCA transition occurs at $t \approx 4\times10^5 \delta_t$. The contact line continues to recede along the hydrophilic stripe until it reaches a hydrophilic-hydrophobic (A-B) boundary, where a sudden jump of the contact line occurs: the contact line quickly crosses the A-B boundary and the hydrophobic stripe. The sudden jump is attributed to a large unbalanced Young's force towards the center of the droplet, arising from the significant deviation between the dynamic contact angle and the equilibrium contact angle of the hydrophobic stripe ($\theta_{B,e}$). The contact line is then pinned on the next hydrophobic-hydrophilic (B-A) boundary, as shown in the third snapshot in Fig.

4(b). Later, the droplet repeats the stick-slip-jump behavior. The second CCR-CCA transition appears at $t \approx 6.1 \times 10^5 \delta_t$ and the second jump of the contact line occurs around $t = 6.6 \times 10^5 \delta_t$. After the second jump (also the last jump) in Fig. 4(a), there are only three stripes wetted by the droplet: the central hydrophilic stripe and two hydrophobic stripes close to it. In the present study, the volume of the droplet which wets three stripes with the contact angle $\theta_{A,e}$ is much larger than the volume of the droplet that wets one stripe with the contact angle $\theta_{B,e}$. Hence it is impossible for the contact line to cross a whole hydrophobic stripe during the last jump. Therefore, after the last jump, the contact line will recede towards the center of the droplet along the remaining part of the hydrophilic stripe, which can be seen from the bottom snapshot in Fig. 4(b).

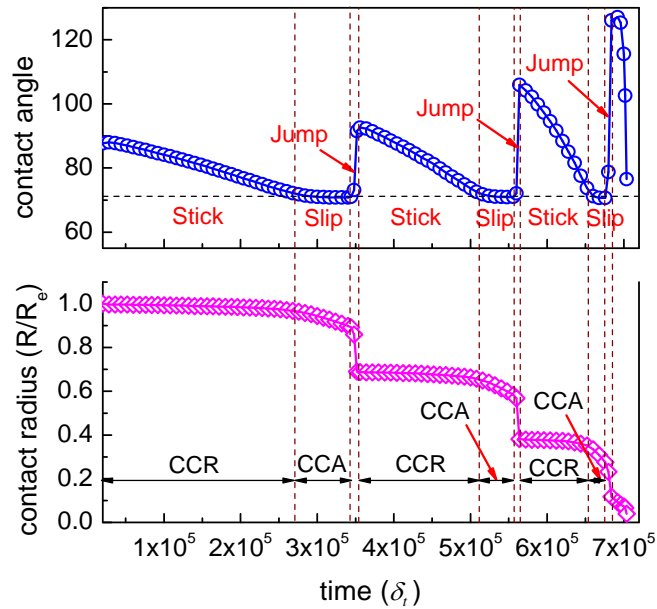

**Fig. 6** Droplet evaporation on the W9 surface. Time evolutions of the contact angle (degrees) and the non-dimensional contact radius.

The numerical results of droplet evaporation on the W17 and W9 surface are displayed in Fig. 5 and Fig. 6, respectively. Similar stick-slip-jump behaviors can be clearly observed in these two cases. The main difference may lie in the repeat times of the stick-slip-jump behavior. From Fig. 6 it can be seen that the stick-slip-jump behavior is repeated three times during the evaporation on the W9 surface while there is only once on the W17 surface as shown in Fig. 5. This difference is expected since the number of the stripes covered by the droplet at the beginning of the evaporation is different in these cases (see Fig. 2). Furthermore, from Fig. 6 we can see that the change of the contact angle is different in each jump. For a later jump, the change of the contact angle is larger than that in an earlier one. This is because the stripe width is gradually comparable to the droplet size, which decreases during the evaporation. Obviously, if the droplet is large enough, reducing the contact radius by a stripe width will have a negligible effect on the contact angle. The aforementioned features, such as the stick-slip-jump behavior and a larger change in the contact angle during a later jump, are consistent with the theoretical prediction in Ref. [33] and the results of molecular dynamics simulations [23]. However, some differences should also be noticed. From Figs. 4, 5, and 6 we can see that the variation of the contact radius during the "stick" (CCR) mode is very small and rapid changes of the contact angle and the contact radius can be observed in each jump, while in Ref. [23] Zhang *et al*. reported an apparent decrease of the contact radius during the "stick" (CCR) mode and a relatively smooth change during the jump stage, which is probably due to the fact that their study was performed at the nanoscale.

The pinning and depinning mechanism of the contact line is now discussed. The analysis in Section 3.1 has provided a basis for understanding the pinning-depinning phenomenon during evaporation on chemically stripe-patterned surfaces. The local details of the liquid-vapor interfaces as well as the density profiles along the solid surface at $t = 10^5 \delta_t$, $2.8 \times 10^5 \delta_t$, and $3.8 \times 10^5 \delta_t$ are displayed in Fig. 7 for droplet evaporation on the W11 surface. From the figure we can observe that there is a movement of the contact line towards the center of the droplet, although the moving distance is very small as compared with the contact radius. Specifically, the length of the contact line occupied by the B stripe gradually decreases, whereas the length occupied by the A stripe increases, which is actually a dynamic adjustment in response to the decrease of the contact angle. When the contact angle $\theta$ $(\theta > \theta_{A,e})$ approaches $\theta_{A,e}$, $|\cos\theta - \cos\theta_{B,e}|$ increases and $|\cos\theta - \cos\theta_{A,e}|$ decreases. Hence the unbalanced Young's force (per unit length) towards the center of the droplet will increase and the unbalanced Young's force towards outside will decrease. To achieve a dynamic equilibrium, the length ratio $l_B$ should decrease while the length ratio $l_A$ should increase, as demonstrated by Fig. 7.

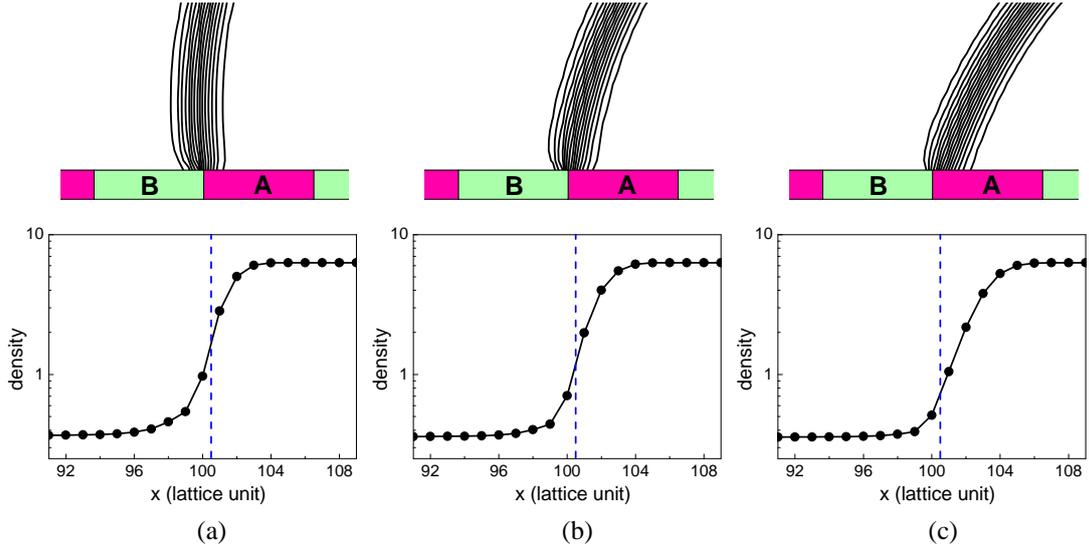

**Fig. 7** Droplet evaporation on the W11 surface. Top: local details of the liquid-vapor interfaces ($0.5 \leq \rho \leq 5.5$) at (a) $t = 10^5 \delta_t$, (b) $t = 2.8 \times 10^5 \delta_t$, and (c) $t = 3.8 \times 10^5 \delta_t$; bottom: the density profile along the solid surface within $x \in [91, 109]$. The B-A boundary (dashed line) is located at $x = 100.5$.

When the contact angle further decreases, the length ratio $l_B$ is close to zero. At a certain instant, the depinning phenomenon occurs. The evaporation goes to the CCA mode and the contact line starts to recede along the hydrophilic stripe. Without thermal effects, the receding contact angle is theoretically equal to the equilibrium contact angle on the pure A surface [33]. In Fig. 4(a), Fig. 5(a), and Fig. 6, we have used a horizontal dashed line to denote the receding contact angle during the CCA (slip) mode. It can be found that the receding contact angles on the three surfaces are basically the same, i.e., $\theta_R \approx 71°$. Since $\theta_R$ is larger than $\theta_{A,e} \approx 63.3°$, an unbalanced Young's force towards outside would appear when the contact line is completely located on an A stripe. In terms of force balance, there should be a local force (or force component) towards the center of the droplet. Otherwise, it is impossible for the contact line to recede along the A stripe with $\theta_R > \theta_{A,e}$.

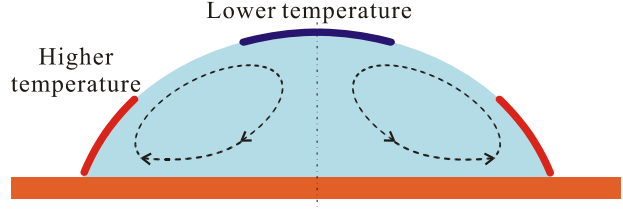

**Fig. 8** A schematic diagram of the Marangoni effect.

We find that the aforementioned phenomenon is attributed to the Marangoni stress on the liquid-vapor interface, which arises from the thermal dependence of the surface tension, namely $d\gamma/dT$ [57, 58], where $\gamma$ is the liquid-vapor surface tension. In some of the previous numerical studies, such as the molecular dynamics study of Wang and Wu [21], the Marangoni effect was not taken into account since the evaporation was triggered by removing the liquid molecules from the droplet surface. For most fluids, the surface tension decreases with increasing temperature, i.e., $d\gamma/dT < 0$. Therefore the surface tension is lower in the place where the temperature is higher and vice versa. The presence of such surface tension gradients results in a surface flow that is directed from regions of lower surface tension (warmer parts) to regions of higher surface tension (colder parts) and a convective flow within the droplet (see Fig. 8), which is usually referred to as Marangoni flow. Figure 9 displays the streamlines at $t = 150000\delta_t$ in the evaporation on the W9 surface. From the figure we can clearly observe the surface flow at the liquid-vapor interface and the convective flow within the droplet, which demonstrates the existence of the Marangoni effect in our simulations. We have evaluated the dependence of the surface tension on the temperature so as to obtain $d\gamma/dT$. Quantitatively, in the above simulations $d\gamma/dT \approx -0.85/T_c$, where $T_c$ is the critical temperature.

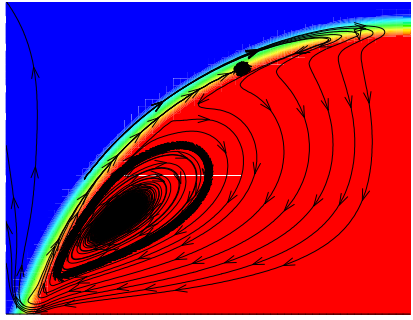

**Fig. 9** The streamlines at $t = 150000\delta_t$ in the evaporation on the W9 surface.

In the literature there have been many studies that focused on the "coffee ring stain" caused by the Marangoni effect. However, very few studies were conducted about the influence of the Marangoni stress on the shape of the droplet or the contact angle during evaporation. Recently, by investigating evaporating sessile droplet of perfectly wetting liquids with small finite contact angles, Tsoumpas *et al*. [59] experimentally showed that, when the Marangoni effect is strong, the droplet profile appears to be inflated (the droplet height increases) as compared to the classical static shape. The results of our simulations are consistent with the finding of Tsoumpas *et al*. Since $\theta_R > \theta_{A,e}$, it is obvious that the droplet height during the slip (CCA) mode is larger than that of a static droplet (with the same volume) on the pure A surface without evaporation. To further illustrate the influence of the Marangoni stress, another case is considered, in which the surface tension and $|d\gamma/dT|$ are reduced by about 30% using the treatment proposed in Ref. [60]. The results (the dynamic contact angle) of the new case on the W9 surface are presented in Fig. 10. For comparison, the results of the original case are also shown there. Owing to the decrease of the surface tension, the evaporation is speeded up in the new case. Moreover, from the figure we can see that in the new case the receding contact angle during the slip (CCA) mode is lowered, $\theta_R \approx 67.5°$, approaching the equilibrium contact angle on the pure A surface, which is expected since the influence of the Marangoni stress has been reduced by decreasing $|d\gamma/dT|$. In the absence of the Marangoni effect ($d\gamma/dT = 0$), $\theta_R$ is theoretically equal to $\theta_{A,e}$ [33].

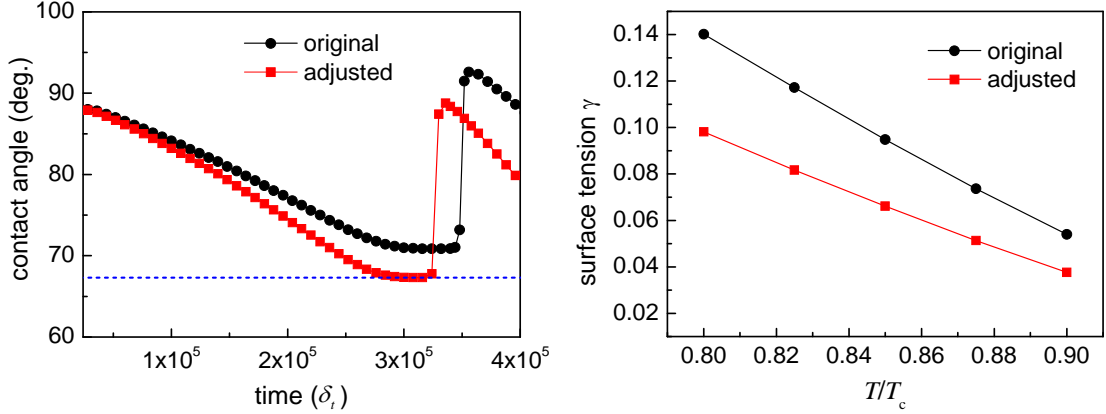

**Fig. 10** (Left) Comparison of the contact angle during the evaporation on the W9 surface. Here "adjusted" represents the case in which the surface tension and $|d\gamma/dT|$ have been reduced by about 30%, while "original" denotes the case without adjustment. (Right) Variations of the surface tension against the temperature.

## 4. Summary

In this paper, we have numerically investigated the pinning and depinning mechanism of the contact line during droplet evaporation on chemically stripe-patterned surfaces using a hybrid thermal multiphase LB model. The width of the hydrophilic stripe is equal to that of the hydrophobic stripe. Three different surfaces were considered: W9, W11, and W17 surfaces, with the stripe width being set to 9, 11, and 17 (l.u.), respectively. First, the equilibrium states of droplets on chemically stripe-patterned surfaces without evaporation were investigated. In the context of diffuse interface modeling, a local force balance has been introduced based on the unbalanced Young's force, which yields an equation like the Cassie equation for evaluating the equilibrium contact angles on chemically stripe-patterned surfaces. Specifically, it is shown that the equilibrium contact angles can be evaluated according to the length of the contact line occupied by each component.

Later, we studied the dynamics of evaporating droplets on the W9, W11, and W17 surfaces. Numerical results show that the droplets evaporate in a "stick-slip-jump" fashion on the three surfaces.

The repeat times of the stick-slip-jump behavior are determined by the number of the stripes covered by the droplet. The sudden jump of the contact line is caused by a large unbalanced Young's force towards the center of the droplet, arising from the deviation between the dynamic contact angle and the equilibrium contact angle of the hydrophobic stripe.

During the stick mode, a very slow movement of the contact line is observed, which demonstrates that the pinning of the contact line during droplet evaporation on chemically stripe-patterned surfaces is a dynamic pinning process and the dynamic equilibrium is achieved by the self-adjustment of the contact lines occupied by each component. Furthermore, the depinning of the contact line is found to be affected by the Marangoni stress, which results from the dependence of the surface tension on the temperature. When $\mathrm{d}\gamma/\mathrm{d}T < 0$, the Marangoni stress promotes the depinning of the contact line. As a result, the receding contact angle during the slip mode (also the depinning contact angle) is larger than the equilibrium contact angle of the hydrophilic stripe. The limitations of the present study are also pointed out here. Owing to using a diffuse interface, the present simulations are limited to the cases in which the stripe width is larger than the thickness of the diffuse interface. Moreover, it should be noted that Eq. (9) is derived in a two-dimensional situation. Three-dimensional analyses and simulations will be performed in the future. In addition, possible experimental validation can also be considered, such as droplet evaporation on surfaces patterned with hydrophobic and hydrophilic concentric rings.

## Acknowledgments

This work was supported by the National Natural Science Foundation of China (No. 51506227), the Foundation for the Author of National Excellent Doctoral Dissertation of China (No. 201439), and the Los Alamos National Laboratory's Lab Directed Research & Development Program.

**TOC graphic**

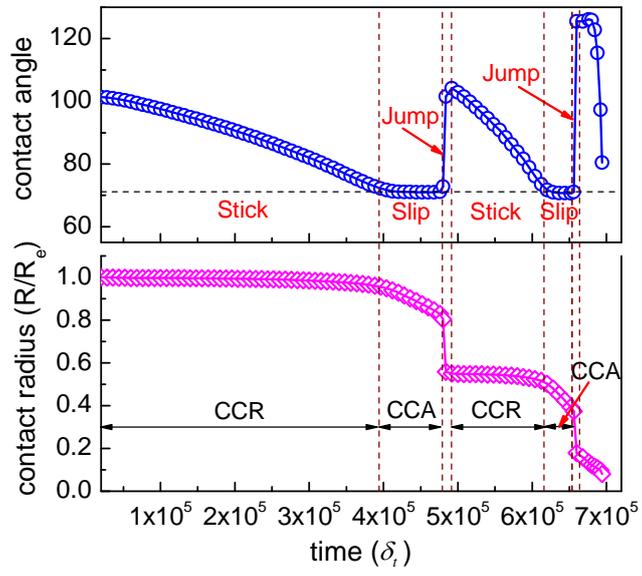